\documentclass[12pt]{article}
\usepackage{a4wide}
\usepackage{amssymb}
\usepackage{graphicx}
\begin{document}
{\renewcommand{\thefootnote}{\fnsymbol{footnote}}
\begin{center}
{\LARGE  Emergent Modified Gravity }\\
\vspace{1.5em}
Martin Bojowald\footnote{e-mail address: {\tt bojowald@psu.edu}}
and Erick I.\ Duque\footnote{e-mail address: {\tt  eqd5272@psu.edu}}
\\
\vspace{0.5em}
Institute for Gravitation and the Cosmos,\\
The Pennsylvania State
University,\\
104 Davey Lab, University Park, PA 16802, USA\\
\vspace{1.5em}
\end{center}
}

\setcounter{footnote}{0}

\begin{abstract}
  A complete canonical formulation of general covariance makes it possible to
  construct new modified theories of gravity that are not of higher-curvature
  form, as shown here in a spherically symmetric setting. The usual uniqueness
  theorems are evaded by using a crucial and novel ingredient, allowing for
  fundamental fields of gravity distinct from an emergent space-time metric
  that provides a geometrical structure to all solutions. As specific
  examples, there are new expansion-shear couplings in cosmological models, a
  form of modified Newtonian dynamics (MOND) can appear in a space-time
  covariant theory without introducing extra fields, and related effects help
  to make effective models of canonical quantum gravity fully consistent with
  general covariance.
\end{abstract}

\section{Introduction}

General relativity and possible modifications or alternatives, such as quantum
gravity corrections, can, to a large extent, be derived from the symmetries of
space-time. In the usual formulation based on space-time tensors and
Riemannian geometry, the gravitational action must be invariant under
coordinate changes and, since it depends on the space-time metric (or
alternative geometrical objects such as tetrads or connections), is therefore
given by a curvature or torsion invariant. In general relativity, the relevant
invariant is proportional to the space-time Ricci scalar plus, perhaps, a
cosmological constant. Einstein's field equation then follows from the
variational principle. If higher-order derivatives are included in an
effective action that may describe certain quantum effects, higher-curvature
actions are obtained in which the Ricci scalar $R$ may be replaced by a
non-linear function $f(R)$, and invariants constructed from multiple factors
of the full Riemann or torsion tensors are also possible.

For certain aspects of quantum gravity, a canonical formulation is useful
because the underlying phase-space structure allows one to look for quantum
representations on a Hibert space. Canonical formulations distinguish between
configuration degrees of freedom and their velocities, which are turned into
momenta as independent phase-space variables. At this basic step, therefore,
symmetries such as space-time covariance may not be explicitly realized, but
once all dynamical equations have been solved, the solutions are equivalent to
those of general relativity and are therefore covariant. The dynamical
equations contain, as usual, equations of motion of second order in time
derivatives of the metric (or of first order of the configuration variables
and momenta), but also constraints that are of at most first order in time
derivatives (or contain only spatial derivatives of the canonical fields). Both
types of equations follow from specific components of the Einstein tensor, and
they can be independently derived in the canonical formulation.

Early, classic work \cite{DiracHamGR,Katz,ADM,Regained} has shown that the
constraints are the relevant part of these equations when it comes to
symmetries: When the constraints are solved, a condition referred to in this
context as going ``on-shell,'' general covariance can be recovered in the
canonical theory from flow equations generated by the constraints. The
constraints generate evolution equations as well as gauge transformations by
their Hamiltonian vector fields, which on shell are equivalent to space-time
coordinate changes. However, the tensorial and canonical formulations are not
equivalent off-shell, which may be relevant in particular for an understanding
of possible quantum effects that could modify the classical constraint
functions in different ways.

Here, we analyze this question in the simpler context of modified gravity and
in a reduction to spherically symmetric models. As we will demonstrate, novel
covariant theories of modified gravity are then possible, with interesting new
phenomena. For instance, there are new types of dynamical
  signature change \cite{EmergentSig}, and previously constructed models of
  non-singular black holes \cite{SphSymmEff,SphSymmEff2} can be rederived as
  special cases.  These results help to clarify possible modified-gravity
effects implied by canonical quantum gravity. The resulting theories may be of
interest also for purely classical questions, for instance by providing new
covariant equations for phenomenological applications that cannot be obtained
from traditional versions of modified gravity.

Physically, new theories of modified gravity are possible in a canonical
formulation because setting up a canonical theory of gravity requires weaker
assumptions than what is used to specify an action. Any tensorial formulation
of gravity by an action principle has to start with a fundamental space-time
tensor, usually the metric, on which the Lagrangian depends and which defines
the integration measure for the action. A canonical formulation of gravity, by
contrast, can be given with fewer assumptions because it only requires
suitable spatial tensors with a phase-space structure, and only spatial
integrations in the Hamiltonian and diffeomorphism constraints. Unlike in a
tensorial formulation, in which general covariance is built into the formalism
by making use of the tensor transformation law in space-time, general
covariance in a canonical formulation is a derived concept that must be
demonstrated by an analysis of Poisson brackets and gauge flows of the
constraints. While this property makes it harder to construct covariant
theories of gravity in purely canonical form, it also provides an opening to
new theories of modified gravity because a canonical formulation starts with
weaker requirements on the fundamental fields.

In particular, we will see that it is possible to relax the usual identity
between the fundamental canonical configuration field and the induced spatial
metric of the corresponding space-time geometry. This identity is realized in
all higher-curvature formulations of modified gravity, but it is not necessary
in a broader setting of modified canonical gravity. In these new theories, the
space-time metric, or even its spatial part, is not fundamental but rather
emergent. Even if it does not agree with a canonical configuration variable,
the emergent spatial metric $q_{ab}$ is uniquely determined by the structure
function of the Poisson bracket of two Hamiltonian constraints.

Moreover, the canonical realization of general covariance works by
reconstructing space-time transformations from deformations of spatial
hypersurfaces in normal and tangential directions. In a canonical formulation
of a theory defined by an action principle directly for the space-time metric
or tetrad, such as general relativity or some higher-curvature theory, the
normal direction, just as the induced spatial metric on a hypersurface, is
uniquely determined by the fundamental space-time metric. If one tries to
construct consistent gravity theories purely in a canonical setting, the
normal direction is defined more abstractly through the rules by which general
covariance is implemented canonically, related to the algebraic form of
different Poisson brackets of the constraints that define the theory. In the
absence of a fundamental space-time metric, it turns out, this definition of
the normal direction $n^a$ is no longer unique and may allow inequivalent
realizations of a geometrical space-time structure. Each realization results
in a consistent and covariant gravitational theory with an emergent space-time
metric subject to the full set of coordinate transformations, defined by the
usual combination $g_{ab}=q_{ab}-n_an_b$ where both objects on the right are
now emergent, $q_{ab}$ from the structure function and $n_a$ from what has
been specified as the Hamiltonian constraint.  But different realizations
imply inequivalent emergent space-time metrics not mutually related by
coordinate changes. In practice, different choices of normal directions can be
parameterized by replacing the original constraints of canonical general
relativity with suitable linear combinations subject to phase-space dependent
coefficients.

Such linear combinations therefore present a new, previously unrecognized
approach to modified gravity. This outcome is surprising because one usually
thinks of the constraints as generators of some algebraic structure, which
should be invariant under linear combinations of the generators. However,
equipping solutions with a geometrical structure via the emergent spatial
metric and normal direction is an additional ingredient that may well be (and,
as we will show explicitly, is) sensitive to which linear combinations of
algebraic generators are distinguished as the specific Hamiltonian and
diffeomorphism constraint whose gauge transformations correspond to normal and
tangential directions of spatial hypersurfaces.  The well-known structure
function in the Poisson bracket of two Hamiltonian constraints then determines
the emergent space-time metric which, subject to strong consistency
conditions, also depends on which linear combination of the generators is
singled out as the Hamiltonian constraint.

Since the foundation of our new theories of emergent modified gravity depends
crucially on these canonical structures and properties of hypersurface
deformations, we will begin with a detailed review of relevant properties in
the next section. The resulting new spherically symmetric theories, along with
some solutions, will then be evaluated for new physics effects, considering
expansion and shear terms in cosmological models as well as regimes of
intermediate-strength gravity. Some of the new terms could be implied by
various ingredients of quantum gravity, but a purely classical interpretation
is also possible in which novel covariant modifications can be made available
for phenomenological studies. The main motivation of emergent modified gravity
then consists in the observation that the space-time metric need not be one of
the fundamental fields in an action principle while maintaining the symmetry
condition of general covariance. From the perspective of effective field
theory, any new terms made possible by this more general viewpoint should then
be included in physical evaluations, especially in tests of strong-field
gravity.

\section{Hypersurface deformations}

Following \cite{Regained}, the gauge symmetries of the canonical formulation
of general relativity can be identified with hypersurface deformations of
constant-time spatial slices in space-time, operating in normal and tangential
directions. Since a coordinate transformation in general changes the notion of
constant-time slices, hypersurface deformations are related to coordinate
transformations. In practice, indeed, standard methods to analyze solutions of
general relativity may refer to coordinate transformations or to choosing a
slicing of space-time into a collection of spatial hypersurfaces without making much
of a conceptual distinction. However, while a coordinate choice in particular
of time defines a slicing through constant-time hypersurfaces, the deformation
of a hypersurface in its normal direction is, in general, not the same as a
coordinate change in the time direction. For a given geometrical structure of
space-time, it is not difficult to translate these two notions into each
other. But the construction of new gravitational theories is more involved
because a pre-existing space-time structure cannot be taken for granted.

\begin{figure}
\begin{center}
  \includegraphics[width=14cm]{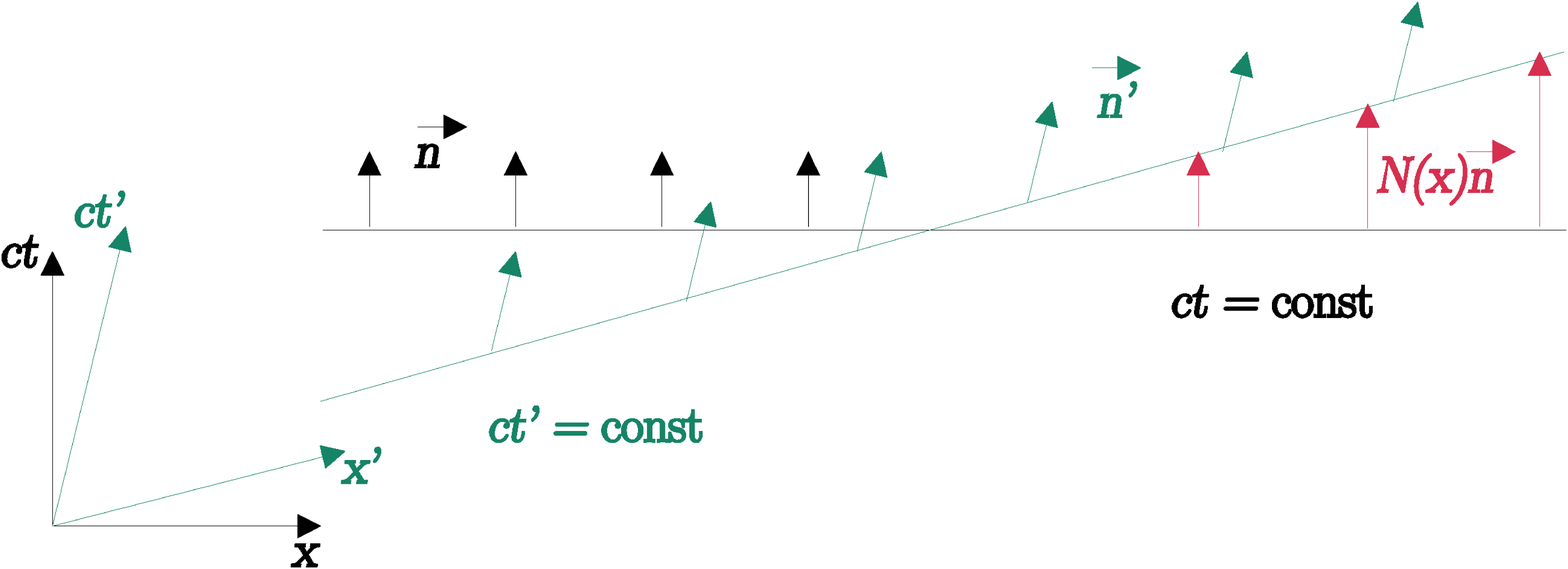}
  \caption{Deformations of constant-time spatial slices in Minkowski
    space-time are equivalent to Poincar\'{e}
    transformations. Spatial slices of constant $t$ and $t'$, related by a
    Lorentz boost, define different hyperplanes in Minkowski
    space-time. Their unit normals $\vec{n}$ and $\vec{n}'$, respectively,
    are defined by the indefinite Minkowski scalar product. A boost can be
    described geometrically by a deformation of the initial slice by a
    position-dependent deformation of length $N(x)$ in the normal direction,
    introducing the lapse function $N(x)$. Similarly, spatial translations
    and rotations are determined by a shift vector field $\vec{w}$ tangential
    to a spatial hypersurface. Combining a pair of hypersurface deformations
    in the two possible orderings implies a commutator that is equivalent to
    the algebraic Lie bracket in the Poincar\'{e} algebra. The example shown
    here represents a geometrical version of the time deformation $T(N)$,
    which leads to the commutator (\ref{TT}) for linear $N(x)$ and a
    Euclidean spatial metric in the case of special relativity. \label{fig:LorentzMink}}
\end{center}
\end{figure}

\begin{figure}
\begin{center}
  \includegraphics[width=8cm]{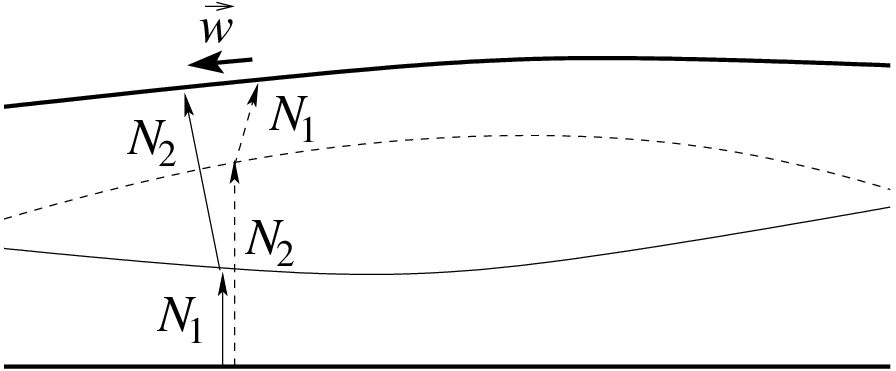}
  \caption{Normal deformations of spatial hypersurfaces in curved
    space-time. Applying a pair of normal deformations with lapse functions
    $N_1$ and $N_2$, respectively, in the two different orderings implies a
    non-zero spatial deformation tangential to the final hypersurface of the
    form (\ref{TT}). The normal angle is shown based on local Minkowski
    geometry, following Fig.~\ref{fig:LorentzMink}. In general relativity, the
    functions $N(x)$ need not be linear, and induced spatial metrics on
    hypersurfaces may be curved. The displacement vector $\vec{w}$ then
    depends on the spatial metric through the gradients of $N_1$ and $N_2$ in
    (\ref{TT}). \label{fig:SurfaceDefMink}} 
\end{center}
\end{figure}

In the simpler case of special relativity, in which the Minkowski space-time
metric is constant in Cartesian coordinates and only linear transformations
are considered, the symmetries of the Poincar\'{e} algebra can be shown to be
equivalent to hypersurface deformations, as illustrated in
Fig.~\ref{fig:LorentzMink}. But the relationship between hypersurface
deformations and space-time transformations is more complicated if one goes
beyond this setting by allowing for non-Cartesian coordinates or space-time
curvature.  At first sight, the transition to general relativity looks simple:
Hypersurface deformations can easily be generalized to non-planar spatial
hypersurfaces in curved space-time and their deformations along normal and
tangential directions. As illustrated in Fig.~\ref{fig:SurfaceDefMink}, given
a background space-time metric in which these deformations take place, one can
compute geometrical commutators of pairs of such deformations. Using
$S(\vec{w})$ to denote an infinitesimal spatial deformation by a shift vector
field $\vec{w}$ tangential to a hypersurface, and $T(N)$ to denote a timelike
deformation along the unit normal by a lapse function $N$, the well-known
result is given by the equations \cite{DiracHamGR,Katz,ADM}
\begin{eqnarray}
 [S(\vec{w}_1),S(\vec{w}_2)]&=& S({\cal L}_{\vec{w}_1}\vec{w}_2)\label{S}\\
{} [T(N),S(\vec{w})] &=& -T({\cal L}_{\vec{w}}N)\\
{} [T(N_1),T(N_2)] &=& S(N_1\nabla N_2-N_2\nabla N_1) \label{TT}
\end{eqnarray}
where ${\cal L}_{\vec{w}}$ is the standard directional or Lie derivative along
the vector field $\vec{w}$. 

While the last equation, (\ref{TT}), looks quite simple, it complicates the
underlying mathematical structure because, unlike the Lie derivatives in the
first two equations, the gradient $\nabla N$ of a function $N$ requires a
metric. Such a metric is available, given by the spatial metric
\begin{equation} \label{qg}
  q_{ab}=g_{ab}+n_an_b
\end{equation}
induced by the space-time metric $g_{ab}$ on an embedded hypersurface with
unit normal $n^a$. But its presence in (\ref{TT}) means that the commutator
not only depends on the generating functions $N$ and $\vec{w}$ that define
hypersurface deformations, but also on the geometry on the hypersurface. In
the physics literature, such a bracket is usually referred to as ``open'' or
one with ``structure functions.'' To close the brackets, we have to be able to
apply commutators multiple times, which, since the spatial metric appears in
$S$ on the right-hand side of (\ref{TT}), requires that $N$ and $\vec{w}$
should be allowed to depend on the metric, in addition to their dependence on
coordinates. With this additional dependence, the set of generators is larger
than what the four space-time directions would initially
indicate. (Mathematically, the metric dependence can be formulated by using
the notion of algebroids, in which (\ref{S})-(\ref{TT}) are realized as
brackets on sections of a fiber bundle over a suitable space of
metrics. However, as shown recently \cite{ConsRinehart} in an application of
general mathematical results to the case of hypersurface deformations, the
corresponding bracket, placed in the language of BRST/BFV gauge generators, cannot
be Lie but is $L_{\infty}$.)

For our new results, the appearance of the spatial metric in the structure
function of an algebraic bracket turns out to be a crucial ingredient. If we
start from scratch in order to define a canonical theory by finding expressions for
$S(\vec{w})$ and $T(N)$ in agreement with (\ref{S})--(\ref{TT}), we will have
to assume a certain phase space on which these expressions are defined, using
Poisson brackets to compute a realization of the required
commutators. However, since the spatial metric appears in the structure
function of any set of generators that has the form of (\ref{S})--(\ref{TT}),
we do not have to make assumptions about which phase-space function can play
the role of a spatial metric. The geometrical structure of solutions, if it
exists given consistency conditions described below, is therefore a derived
concept in canonical gravity and need not be presupposed. This conclusion
would not be possible if the generators had formed a Lie algebra with structure
constants.

In practice, we usually do not try to invent completely new constraints that
could have a chance of obeying (\ref{S})--(\ref{TT}). We rather start with the
known classical constraints of a gravity theory in canonical form, and try to
modify them so as to preserve as much of the structure of their Poisson
brackets. This condition together with other requirements of general
covariance is very restrictive, as shown for instance in several discussions
related to models of loop quantum gravity \cite{SphSymmCov,GowdyCov,EffLine}
which, as we will describe below, did not even include all the required
covariance conditions. In these examples, the main modification consisted in a
non-polynomial dependence of the Hamiltonian constraint or the generator
$T(N)$ on the classical expression for extrinsic curvature. While partial
realizations of such modified covariant models were possible in vacuum
spherical symmetry, an extension to models with local degrees of freedom, such
as spherical symmetry with scalar matter or polarized Gowdy models, turned out
to be difficult.

A different way of modifying canonical gravity has been implicitly suggested
by recent work on spherically symmetric models
\cite{SphSymmMatter,SphSymmMatter2,SphSymmEff,SphSymmEff2}, building on the
older \cite{Absorb} and \cite{EffLine}. The constructions in
\cite{SphSymmEff,SphSymmEff2} used different ingredients not considered in
what follows, such as non-bijective canonical transformations, but the crucial
step, as we will demonstrate, was an apparently innocuous application of
linear combinations of the original constraints with phase-space dependent
coefficients. Such coefficients, or equivalently a phase-space dependence of
$N$ or $\vec{w}$ in $T(N)$ and $S(\vec{w})$, changes the brackets
(\ref{S})--(\ref{TT}).  For instance, if $N_1$ and $N_2$ depend on the
phase-space degree of freedom $q$ that classically represents the spatial
metric, the Poisson bracket $\{T(N_1),T(N_2)\}$ contains terms of the form
$T(\partial N/\partial q)$ in contrast to (\ref{TT}). Such terms vanish
on-shell, but they are relevant for the off-shell brackets and possible
space-time geometries that could be reconstructed from them. If there is a
contribution from $T$ to the right-hand side of (\ref{TT}), the bracket does
not have the form required for hypersurface deformations, even if it is still
closed (and therefore anomaly-free in the sense of gauge
transformations). However, it may be possible to remove such terms by further
modifications of the generators, in particular replacing the $T(N)$ that one
identifies with the Hamiltonian constraint with a linear combination of the
original $T(N)$ and $S(\vec{w})$, both with phase-space dependent
coefficients. The form (\ref{TT}) may then be recovered, although in general
the structure function does not equal the original phase-space degree of
freedom $q$ (or its inverse).

Crucially, unlike in an algebraic bracket in which $T(N)$ is just one of the
generators, the canonical reconstruction of a space-time geometry depends on
which expression $T(N)$ one considers to represent normal
deformations. Singling out a specific $T(N)$ with phase-space dependent $N$ as
the Hamiltonian constraint, or a specific linear combination of both $T(N)$
and $S(\vec{w})$ with phase-space dependent coefficients, {\em defines} the
normal direction of hypersurfaces. The choice can therefore have an effect on
the resulting reconstructed space-time geometry. Similarly, if the structure
function in (\ref{TT}) is modified, the geometry of hypersurface deformations
implies that the inverse of the new structure function has to be identified
with the spatial metric of a reconstructed space-time line element.  Physical
evaluations of versions of $T(N)$ with different dependencies of $N$ on the
spatial metric are not guaranteed to be equivalent to one another because they
imply different space-time metrics from the canonical ingredients, given by
the definition of a normal direction through the choice of $T(N)$ and the
spatial metric derived from the structure function in (\ref{TT}). Since the
space-time line element is derived in this situation and does not agree with
the original phase-space degrees of freedom, we refer to it as an {\em
  emergent line element}.

This new and unexpected feature makes it possible to look for novel theories
of modified gravity that are not of the form of
  higher-curvature or other action-based theories. Instead of starting with a
  space-time tensorial object such as a metric or connection and then looking
  for invariant action functionals depending on these basic fields, emergent
  modified gravity starts with a phase space suitable for gravitational fields
  and derives a compatible object for space-time geometry as well as its
  dynamics. This object may depend non-trivially on the basic fields,
  generalizing the usual construction method based on action principles. In
what follows, we confirm this expectation by an explicit construction in the
setting of spherically symmetric gravity, highlighting consistency conditions
that make it possible to construct an emergent line element invariant under
the full set of coordinate transformations. Our examples will reveal new
effects in cosmological models as well as modified Newtonian gravity in a
covariant theory.

\section{Covariance and emergent line elements}

Spherically symmetric gravity can be described by line elements
\begin{equation} \label{ds}
    {\rm d}s^2= -N(t,x)^2{\rm d}t^2+ \frac{e_2(t,x)^2}{e_1(t,x)} ({\rm
      d}x+w(t,x){\rm d}t)^2+ e_1(t,x) {\rm d}\Omega^2
\end{equation}
where $e_1>0$, $e_2$, $N$ and $w$ are functions of the time coordinate $t$ and
the radial coordinate $x$ but do not depend on the polar angles $\vartheta$
and $\varphi$ that appear in
${\rm d}\Omega^2={\rm d}\vartheta^2+\sin^2\vartheta{\rm d}\varphi^2$. The
momenta $k_1$ and $k_2$, canonically conjugate to $e_1$ and $e_2$,
respectively, also depend only on $t$ and $x$. Explicit phase-space
expressions classically realizing $T(N)$ and $S(w)$ as functions of $e_i$ and
$k_i$ are given by \cite{SphSymmHam}
\begin{equation} \label{H}
   T [N] = \int {\rm d} x\ N \bigg( 
    - \frac{e_2 k_2^2}{2 \sqrt{e_1}}
    - 2 k_2 \sqrt{e_1} k_1 
    - \frac{\sqrt{e_1} e_1' e_2'}{2 e_2^2} 
    + \frac{\sqrt{e_1} e_1''}{2 e_2} +\frac{(e_1')^2}{8 \sqrt{e_1} e_2}
    - \frac{e_2}{2 \sqrt{e_1}}
    \bigg)
\end{equation}
and
\begin{equation} \label{D}
    S [w] = \int {\rm d} x\ w \left( k_2' e_2 - k_1 e_1' \right)
  \end{equation}
(using units in which Newton's constant equals one).
Their Poisson brackets obey (\ref{TT}) provided $N$ and $w$ do not depend on
$e_i$ and $k_i$. Equations of motion generated by these constraints identify
$k_1$ and $k_2$ with suitable components of extrinsic curvature, but this
relationship is in general modified when new expressions for $T[N]$ and $S[w]$
are used.

Motivated by the discussion of the previous section, we now implement a
specific linear transformation of the classical generators $T(N)$ and $S(w)$,
represented by (\ref{H}) and (\ref{D}), by replacing them with $T(N')$ and
$S(w')$ with $N'=\alpha N$ and $w'=\beta N+w$, where $\alpha$ and $\beta$ are
allowed to depend on $(e_i,k_i)$. (Without loss of generality, we will assume
that the original $N$ and $w$ are not phase-space dependent.) In terms of the
old $N$ and $w$, the new combined generator
\begin{equation} \label{linear}
  T(N')+S(w')= [T(\alpha N)+S(\beta N)] +S(w)
\end{equation}
contains an unchanged spatial deformation $S(w)$, but the previous normal
deformation $T(N)$ is replaced by $T'(N)=T(\alpha N)+S(\beta N)$. If it is
possible to reconstruct a generally covariant space-time geometry from the new
normal direction and the resulting structure function, it will be different
from the classical line element defined by (\ref{ds}) even though it will be a
gravitational theory for the same fields $e_1$, $e_2$ and $N$.

In order to make sure that the transformed generators still define a
space-time geometry suitable for a gravitational theory, such that solutions
are generally covariant in the usual sense and can be described by invariant
line elements familiar from Riemannian geometry, we must show that $T(N')$ and
$S(w')$ still have Poisson brackets of the form (\ref{S})--(\ref{TT}). Any
extra terms from phase-space derivatives of $\alpha$ and $\beta$ must
therefore combine in the correct way for the right-hand sides of
(\ref{S})--(\ref{TT}) to appear. If this is the case, there may still be
deviations in the precise coefficients, in particular in (\ref{TT}) where a
different structure function may appear instead of the inverse spatial metric
of (\ref{ds}) used in the gradient.  Therefore, traditional uniqueness results
going back to \cite{DiracHamGR} need not apply, opening up the possibility of
new modified theories of gravity. However, a modified gravity theory of
classical type, described by a line element, must still be generally
covariant. The final task of showing that a modified theory is well-defined
then consists in demonstrating that its gauge transformations, generated by
$T(N')$ and $S(w')$, are equivalent to coordinate changes in an emergent line
element distinct from (but depending on) the fundamental canonical fields.

These tasks, computing the Poisson brackets and checking gauge
transformations, are lengthy exercises, but they are unambiguous because the
expressions for $T$ and $S$ are explicitly known. (See \cite{HigherCov} for
more details.) The result is that the theory may indeed be modified in new
ways, but with restrictions that limit the initial freedom of two functions,
$\alpha$ and $\beta$. The covariance condition requires
\begin{equation} \label{alpha}
    \alpha(e_1,k_2) = \sqrt{1-s\lambda(e_1)^2k_2^2}
\end{equation}
where $s=\pm 1$ is a sign choice and $\lambda$ a free function of $e_1$. (We
are dropping a possible multiplier of the square root that would be allowed to
depend only on $e_1$. The same multiplier would appear in $\beta$ and simply
rescale $T(N')$.)  The second function, $\beta$, is then completely determined
and equals
\begin{equation} \label{beta}
  \beta(e_1,k_2) = \frac{\sqrt{e_1}}{2e_2^2} \frac{\partial
    e_1}{\partial x} \frac{s\lambda(e_1)^2
    k_2}{\sqrt{1-s\lambda(e_1)^2k_2^2}}\,.
\end{equation}

Given these functions, the Poisson-bracket version of (\ref{TT})
produces an expression related to $S(\vec{w})$, but its coefficients do not
exactly equal those in (\ref{TT}), where
$\nabla N=q^{xx}\partial N/\partial x$ and $q^{xx}=e_1/e_2^2$ from (\ref{ds})
in spherical symmetry. It rather equals a similar expression in which the
inverse of
\begin{equation} \label{qeff}
  q^{\rm em}_{xx}= 
  \left(1+ \frac{1}{4e_2^2} \frac{s\lambda(e_1)^2}{1-s\lambda(e_1)^2k_2^2}
    \left(\frac{\partial e_1}{\partial x}\right)^2\right)^{-1}
  \frac{e_2^2}{e_1}
\end{equation}
replaces $q^{xx}$.  We thus derived an emergent metric, (\ref{qeff}), distinct
from the classical field $q_{xx}$, for which a detailed analysis building on
\cite{EffLine} shows that its coordinate transformations are consistent with
gauge transformations generated by $T(N')$ and $S(w')$. (The field $q_{xx}$
can no longer be interpreted as a metric or some other space-time tensor
because its gauge transformations do not imply the tensor transformation
law. There is a single metric in our modified gravity theories, given by
$q^{\rm em}_{xx}$, which are therefore not of bimetric form.) If $\alpha$ and
$\beta$ do not obey (\ref{alpha}) and (\ref{beta}), there is no emergent
metric that obeys this condition, and the theory cannot be covariant or
geometrical.

These results already demonstrate that we have arrived at a new class of
modified gravity theories. They clearly have general relativity as a limit
$\lambda\to0$, and have solutions with nearly standard behavior in regimes in
which $\lambda$ or its multipliers in (\ref{alpha}), (\ref{beta}) and
(\ref{qeff}), such as $k_2$, are small. The fact that the emergent metric of a
covariant line element for solutions of these theories does not agree with the
basic field $q_{xx}=e_2^2/e_1$ in the constraints implies that such theories
cannot follow from a space-time action principle of the usual form: The latter
would require a space-time metric in order to define the integration measure.
But since the correct metric is emergent and not fundamental, the variation of
the metric has to be done with respect to the fundamental fields rather than
the metric to obtain the equations of motion. Since this procedure cannot
provide the expression for the emergent metric which is not known a priori,
there is no fundamental action principle. This observation
  also demonstrates the more general nature of emergent modified gravity
  compared with conventional modifications: In the latter case, the spatial
  metric would, by assumption, be given by the fundamental fields, such as
  $q_{xx}=e_2^2/e_1$ in triad variables. A direct comparison with (\ref{qeff})
  shows that this assumption is compatible only with the case of $\lambda=0$,
  ruling out any modifications of the kind studied here. However, this outcome
  is a consequence only of the assumption that fundamental fields directly
  determine the spatial metric. It is not implied by general covariance of the
  resulting gravitational theory.

If the emergent metric were known, we could use it for the integration measure,
while the equations of motion are obtained by varying with respect to the
fundamental fields. But to the best of our knowledge, the emergent metric
cannot be obtained from purely variational arguments.  The fundamental fields
($e_1$ and $e_2$) are distinct from the emergent metric ($q_{xx}^{\rm em}$),
just as the physical significance of matter, given by the stress-energy
tensor, is distinct from the fundamental fields that define it.  Yet, the
field equations as well as covariance transformations are completely
determined by the constraints and their canonical structure, derivable from
Poisson brackets. These properties are in contrast to standard modified
gravity theories, such as $f(R)$ or TeVeS, which modify the dynamics but not
the structure of space-time because they all assume that the metric is a
fundamental field by necessity.

Any new class of modified gravity theories implies a wealth of applications,
in comparisons with observations or as models of various effects of quantum
gravity. In particular, new effects are not necessarily restricted to large
curvature.  We will present a few examples in what follows.

\section{Applications}

We first briefly mention two implications of our modifications that depend on
characteristic details of the function (\ref{alpha}) and the metric
(\ref{qeff}), and suggest how emergent modified gravity may differ from
standard modifications. These features depend on the sign factor $s$. If
$s=1$, the function (\ref{qeff}) cannot be negative and may directly be used
as an inverse spatial metric. For $s=-1$, however, there may be solutions for
which (\ref{qeff}) changes sign and becomes negative for certain ranges of
$x$. This possibility has been demonstrated explicitly in
\cite{EmergentSig}. The sign then implies signature change, derived
dynamically from a covariant theory of gravity. This possibility cannot appear
if the structure of space-time is presupposed before an action principle is
defined.
  
If $s=1$, the functional form (\ref{alpha}) together with reality conditions
implies that the curvature component $k_2$ must be
bounded. While this outcome may also be obtained for certain
  solutions of higher-curvature theories, such as \cite{NonSingdeSitter}, here
  it is generic for all solutions provided $s=1$.  We can relate this result
to a recent construction \cite{SphSymmEff,SphSymmEff2} of modified spherically
symmetric solutions that turns out to be a special case of our new class. If
we choose $\lambda(e_1)=\lambda$ constant in addition to $s=1$, we can apply
the canonical transformation
\begin{equation} \label{can}
  k_2=\frac{\sin(\lambda k'_2)}{\lambda} \quad,\quad e_2=\frac{e_2'}{\cos(\lambda
    k_2')}
\end{equation}
and obtain $\alpha=\cos(\lambda k_2)$. In this form, the example of covariant
modified gravity had been used in \cite{SphSymmEff,SphSymmEff2} in order to
construct a non-singular black-hole model, but the origin of the modification
remained unclear. Initially it seemed that it may be a consequence of the
non-invertible canonical transformation (\ref{can}), but this possibility had
already been ruled out in \cite{NonCovPol} in a discussion of the earlier
\cite{CovPol}. Our results clarify that the modification of gravity in the
model of \cite{SphSymmEff,SphSymmEff2} is indeed genuine and relies not on a
canonical transformation but rather on a linear, phase-space dependent
transformations of generators that redefines the normal direction of
hypersurface deformations, compared with the classical geometry.

\subsection{General modifications}

More generally, we may try to modify the classical expression of $T(N)$ before
we apply a linear transformation (\ref{linear}).  For instance, as suggested
in models of loop quantum gravity \cite{JR,SphSymmCov}, we could replace the
quadratic dependence on curvature components in $T(N)$ by non-classical
polynomials or even non-polynomial functions, motivated by quantum-gravity
considerations. (In the canonical setting, higher-order expressions of
curvature, given by momentum components, do not imply higher time derivatives,
unlike space-time tensorial higher-curvature actions.) On its own, such a
modification cannot easily be reconciled with general covariance
\cite{EffLine}, but the combination with a suitable linear transformation
(\ref{linear}) turns out to help. Again, this is possible only thanks to the
underlying geometrical behavior of normal directions and emergent metrics,
because from a purely algebraic perspective the linear transformation would
not be able to adjust gauge transformations so as to agree with coordinate
changes of an emergent line element. The main outcome, see again
\cite{HigherCov} for details, is that the emergent metric component
(\ref{qeff}) can be generalized to
\begin{equation} \label{qeff2}
  q^{\rm em}_{xx}(e_1,e_2,k_2)=
    \left(f(e_1)+ \frac{\bar{\lambda}^2}{4e_2^2} 
    \left(\frac{\partial e_1}{\partial x}\right)^2    \right)^{-1}
    \frac{1}{\cos^2 \left(\bar{\lambda} k_2\right)}
     \frac{e_2^2}{e_1}
\end{equation}
with a new free function $f(e_1)$, while $\bar{\lambda}$ is now
constant. (This constant is indirectly related to the previous $\lambda(e_1)$
by a canonical transformation of the phase-space functions
\cite{HigherCov}. If the expression on the right is not
  positive definite, the spatial metric is given by its absolute value and the
  sign determines the signature of space-time.) The emergent spatial metric
(\ref{qeff2}) is realized by a Hamiltonian constraint of the form
\begin{eqnarray}
    T[N] &=& - \frac{1}{2}\int{\rm d}x N \sqrt{e_1} \Bigg( e_2 f_0
    + e_2\left(2\frac{{\rm d} f}{{\rm d} e_1}
    + \frac{f \alpha_2}{ e_1}\right) \frac{\sin^2 \left(\bar{\lambda} k_2\right)}{\bar{\lambda}^2}
 + 4 fk_1  \frac{\sin (2\bar{\lambda} k_2)}{2 \bar{\lambda}}
   \label{Hgen} \\
   &&- \frac{(e_1')^2}{e_2} \left(
    \frac{\alpha_2}{4 e_1} \cos^2 \left( \bar{\lambda} k_2 \right)
    - \frac{k_1}{2e_2} \bar{\lambda} \sin \left(2 \bar{\lambda} k_2 \right)
    \right)
    + \left( \frac{e_1' e_2'}{e_2^2}
    - \frac{e_1''}{e_2} \right) \cos^2 \left( \bar{\lambda} k_2 \right)
    \Bigg)\,. \nonumber 
\end{eqnarray}

Only terms up to second order in spatial derivatives have been
  considered in the derivation. For higher derivative orders, there may be
  additional modified theories for instance of Horndeski type
  \cite{Horndeski,HorndeskiRev,DHOST}. The modifications studied here imply
  second-order field equations, by construction, but they are not included in
  Horndeski-type theories because their emergent space-time metric differs
  from the fundamental fields. Indeed, the crucial Horndeski term of the form
  $(\nabla^{\mu}\phi)(\nabla^{\nu}\phi)\nabla_{\mu}\nabla_{\nu}\phi$ with a
  scalar field $\phi$ does not appear in $T[N]$. In a dilaton interpretation,
  the role of $\phi$ could be played by $e_1$ here, but the terms included do
  not have a sufficient number of derivatives for a dilaton-Horndeski theory
  \cite{DilatonHorndeski}.

The free functions $f_0$ and $\alpha_2$ both depend on
  $e_1$ and play different roles. The function $f_0$ generalizes the classical
  dilaton potential and can also be used to include a cosmological
  constant. The function $\alpha_2$ is restricted by the condition that it
  approaches the value one in the correct low-curvature limit and has no
  classical analog. It is similar to $\alpha$-parameters derived for modified
  canonical theories in \cite{Action}, which can be obtained here in the limit
  $\bar{\lambda}=0$. Since $\alpha_2\not=1$ changes the relative weights
  between the two $k$-terms in $T[N]$, one depending only on $k_2$ and one
  linear in $k_1$, this function may also be interpreted as a modification of
  Ho\v{r}ava--Lifshitz type \cite{Horava}, but one that preserves general
  covariance through the emergent line element.  Like the previous
(\ref{qeff}), the right-hand side of (\ref{qeff2}) is not guaranteed to be
positive for all admissible choices of $f$. We restrict attention here to the
case in which the right-hand side of (\ref{qeff2}) is positive, as required
for a spatial metric. The more general case and the related phenomenon of
signature change are discussed in \cite{HigherCov}.

Given the free functions, the resulting modified theories can
  be analyzed in different ways, depending on physical applications. In order
  to facilitate intuitive interpretations, we first
  continue with the general theory but rewrite the expressions of the emergent
  metric and the Hamiltonian constraint in terms of time derivatives of $e_1$
  and $e_2$ replacing the momenta $k_1$ and $k_2$ upon using equations of
  motion. For simplicity, we assume a vanishing shift function.
  The relevant equations of motion are
  \begin{equation}
    \frac{\dot{e}_1}{N} = 2\sqrt{e_1} \left(f+\frac{\bar{\lambda}^2
        (e_1')^2}{4e_2^2}\right)
    \frac{\sin(2\bar{\lambda}k_2)}{2\bar{\lambda}}
  \end{equation}
  and
  \begin{eqnarray}
    \frac{\dot{e}_2}{N} &=& \sqrt{e_1} \Bigg( 2 e_2 \frac{{\rm d}f}{{\rm d}e_1}
    - \bar{\lambda}^2 \left(\frac{e_1'e_2'}{e_2^2}-\frac{e_1''}{e_2}\right) \Bigg) \frac{\sin(2\bar{\lambda}k_2)}{2\bar{\lambda}}
    \nonumber\\
    &&
    + 2 k_1 \sqrt{e_1} \left( f
    + \bar{\lambda}^2 \frac{(e_1')^2}{4 e_2^2}
    \right) \cos(2\bar{\lambda}k_2)
    + \frac{e_2}{2} \frac{\alpha_2}{e_1} \frac{\dot{e}_1}{N}
    \,.
  \end{eqnarray}
  The first equation, upon using
  \begin{equation}
    \cos^2(\bar{\lambda}k_2)=
    \frac{1}{2}\left(1+\sqrt{1-\sin^2(2\bar{\lambda}k_2)}\right)\,,
  \end{equation}
  can directly be used to write the emergent radial metric as
  \begin{eqnarray}
 q^{\rm em}_{xx}=
    \left( f + \bar{\lambda}^2 \frac{(e_1')^2}{4e_2^2}
    + \sqrt{ \left(f+\frac{\bar{\lambda}^2
        (e_1')^2}{4e_2^2}\right)^2 - \bar{\lambda}^2 \frac{\dot{e}_1^2/N^2}{e_1} }\right)^{-1}
    \frac{2 e_2^2}{e_1}\,.
  \end{eqnarray}
  In this form, the metric depends not only on the gravitational configuration
  variables, $e_1$ and $e_2$, but also on their first-order space and time
  derivatives.

The coefficients in $T[N]$ are then given in terms of
  \begin{equation}
    \frac{\sin(2\bar{\lambda}k_2)}{2\bar{\lambda}}=
    \frac{\dot{e}_1/N}{2\sqrt{e_1}(f+\frac{1}{4}\bar{\lambda}^2(e_1')^2e_2^{-2})}
  \end{equation}
  and
  \begin{equation}
    \cos(2\bar{\lambda}k_2) = \frac{\sqrt{e_1 \left(f+\frac{1}{4}\bar{\lambda}^2(e_1')^2e_2^{-2}\right)^2 - \bar{\lambda}^2 \dot{e}_1^2/N^2}}{\sqrt{e_1}(f+\frac{1}{4}\bar{\lambda}^2(e_1')^2e_2^{-2})}
\end{equation}
as well as $k_1$ derived from $\dot{e}_2$. The resulting expressions are long,
but it is instructive to consider the simpler leading-order corrections in
$\bar{\lambda}^2$. For small $\bar{\lambda}$, we have
\begin{equation}
  \frac{\sin(2\bar{\lambda}k_2)}{2\bar{\lambda}}\approx
  \frac{\dot{e}_1}{2\sqrt{e_1}Nf}
  \left(1-\frac{(e_1')^2}{4e_2^2f}\bar{\lambda}^2\right)
\end{equation}
and hence
\begin{equation}
  \cos(2\bar{\lambda}k_2) \approx
  1-\frac{\dot{e}_1^2}{2e_1N^2f^2}\bar{\lambda}^2
\end{equation}
\begin{eqnarray}
  \cos(\bar{\lambda}k_2) &\approx&
  1-\frac{\bar{\lambda}^2 \dot{e}_1^2}{8 e_1 N^2 f^2}
  \\
  \frac{\sin(\bar{\lambda}k_2)}{\bar{\lambda}} &\approx&
  \frac{\dot{e}_1}{2 \sqrt{e_1} N f} \left( 1
  + \bar{\lambda}^2 \left( \frac{\dot{e}_1^2}{8 e_1 N^2 f^2} - \frac{(e_1')^2}{4e_2^2f} \right) \right)
  \\
\end{eqnarray}
and the emergent radial metric
  \begin{equation} \label{qeffpert}
 q^{\rm em}_{xx} =
\left(1-\frac{1}{4}\bar{\lambda}^2f^{-1}\left(\frac{(e_1')^2}{e_2^2}-
    \frac{\dot{e}_1^2}{e_1N^2f}\right)\right) 
     \frac{e_2^2}{fe_1}\,.
   \end{equation}
   In this form, the leading correction to the radial metric can be written as
   \begin{equation}
     -\frac{1}{4e_1f} g_f^{ab}(\partial_ae_1)(\partial_be_1)
   \end{equation}
   with a classical-type 2-dimensional metric $g_f$ in which $N$ is replaced
   by $\sqrt{f}N$, such that it implies a formal line element
   \begin{equation} \label{qf}
     \not{\!\rm d}s^2=g_{f,ab}{\rm d}x^a{\rm d}x^b= -fN^2{\rm d}t^2+
     \frac{e_2^2}{e_1} {\rm d}x^2\,.
   \end{equation}
   (The notation $\not{\!\!\rm d}$ indicates that this formal line element does
   not obey a covariance condition. It is merely used to summarize common
   coefficients in the object $g_f$.)
 
From the 2-dimensional perspective, $e_1$ is a dilaton field that
   provides derivative terms to the emergent metric.
   From the equation of motion for $\dot{e}_2$, we obtain
   \begin{eqnarray}
 && \frac{\dot{e}_1 e_2}{N} \frac{{\rm d} \log f}{{\rm d} e_1}
+ \frac{\dot{e}_1 e_2 \alpha_2}{2 N e_1} - \frac{\dot{e}_2}{N}
    + \frac{\bar{\lambda}^2}{8 e_1^2 e_2^2 f^3 N^3}
    \Bigg(
    - 4 \dot{e}_1 f^2 N^2 e_1^2 e_1' e_2'
    \nonumber\\
    &&
    e_1 f N^2 (e_1')^2 \left(-4 e_1 \dot{e}_1 e_2 \frac{{\rm d} f}{{\rm d} e_1} + 2 e_1 \dot{e}_2 f-\alpha_2 \dot{e}_1 e_2 f\right)
    \nonumber\\
    &&
    + 2 \dot{e}_1 e_2 \left( \dot{e}_1 e_2 \left(2 e_1 \dot{e}_1 e_2 \frac{{\rm d} f}{{\rm d} e_1}
    - 2 e_1 \dot{e}_2 f + \alpha_2 \dot{e}_1 e_2 f\right)
    + 2 e_1^2 f^2 N^2 e_1'' \right)
    \Bigg)
    \nonumber\\
    &=&
    - 2 k_1 \sqrt{e_1} f 
   \end{eqnarray}
   which determines $k_1$ as a function of $e_1$, $e_2$ and their
   derivatives. Together with the trigonometric expressions for $k_2$ in terms
   of these variables, we obtain the Hamiltonian
   \begin{eqnarray}
    T[N] &=& - \frac{1}{2}\int{\rm d}x N \sqrt{e_1} \Bigg[ e_2 f_0
    - \frac{(e_1')^2}{e_2} \frac{\alpha_2}{4 e_1}
    + \frac{e_1' e_2'}{e_2^2}
    - \frac{e_1''}{e_2}
    \nonumber\\
    &&\qquad
    + \frac{\dot{e}_1}{4 e_1 N^2 f} \left( 4 \dot{e}_2 - \frac{\dot{e}_1}{e_1} e_2 \alpha_2
    - 2 \frac{\dot{e}_1}{e_1} e_2 \frac{{\rm d} \log f}{{\rm d} \log e_1} \right)
    \nonumber\\
  &&
    + \frac{\bar{\lambda}^2 \dot{e}_1}{16 N^4 f^4} \Bigg( f \frac{\dot{e}_1^2}{e_1^2} \left(8 \dot{e}_2 - 3 \alpha_2\frac{\dot{e}_1}{e_1} e_2\right)
    + \frac{\dot{e}_1}{e_1} e_2 \left(4 f N^2 \frac{(e_1')^2}{e_2^2}-6 \frac{\dot{e}_1^2}{e_1} \right) \frac{{\rm d} f}{{\rm d} e_1}
    \nonumber\\
   &&\qquad
    + f^2 N^2 \left(4 \frac{\dot{e}_1}{e_1} \frac{e_1' e_2'}{e_2^2} + \frac{(e_1')^2}{e_1 e_2^2} \left(\alpha_2 \frac{\dot{e}_1}{e_1} e_2 - 4 \dot{e}_2\right)
    - 4 \frac{\dot{e}_1}{e_1} \frac{e_1''}{e_2}\right)
    \Bigg)
\Bigg]\,. 
\end{eqnarray}
 The corresponding Lagrangian equals
 \begin{eqnarray}
  L[N]&=&\int{\rm d}x (-k_1\dot{e}_1-k_2\dot{e}_2)-T[N]\nonumber\\
  &=&
    - \frac{1}{2} \int{\rm d}x N \sqrt{e_1} \Bigg[
    \frac{1}{4}\alpha_2\frac{e_2}{e_1^2} g_f^{ab}(\partial_ae_1)(\partial_be_1) 
    - \frac{1}{e_1} g_f^{ab}(\partial_ae_1)(\partial_be_2) 
    \nonumber\\
&&\quad
    - \frac{e_2\dot{e}_1^2}{2 e_1 N^2 f^2}
      \frac{{\rm d} f}{{\rm d} e_1}    + \frac{e_1''}{e_2}- e_2 f_0 
    \nonumber\\
   &&\quad
    + \frac{\bar{\lambda}^2}{16 f} \Bigg(
\frac{\dot{e}_1(\alpha_2 \dot{e}_1 e_2-4\dot{e}_2e_1)}{e_1^3fN^2}
          g_f^{ab}(\partial_ae_1)(\partial_be_1) +
          4\frac{\dot{e}_1^2}{e_1^2fN^2}g_f^{ab}(\partial_ae_1)(\partial_be_2)
          \nonumber\\ 
   &&\quad
          +\frac{8}{3e_1^2} \frac{\dot{e}_1^3\dot{e}_2 }{f^2N^4} 
    + \frac{\dot{e}_1^2}{e_1fN^2} e_2 \left( 4 \frac{(e_1')^2}{e_2^2} - 2
         \frac{\dot{e}_1^2}{e_1fN^2}\right) \frac{{\rm d} \log f}{{\rm d} e_1}
    - 4 \frac{\dot{e}_1^2}{e_1fN^2} \frac{e_1''}{e_2}
    \Bigg)
   \Bigg]
   \,.
 \end{eqnarray}

Also here, the formal metric coefficients $g_f$ from (\ref{qf}) can be used
 to combine some but not all of the spatial and temporal derivative terms. As
 a characteristic of emergent modified gravity, the actual space-time metric
 $g_{\rm em}$ or its spatial part $q_{\rm em}$ are not directly recognizable
 as a coefficient in the Lagrangian. The extension of the Lagrangian to an
 action with a well-defined space-time integration is therefore not
 obvious. In this sense, emergent modified gravity is different from
 Horndeski-type theories \cite{Horndeski,HorndeskiRev,DHOST}, which also lead
 to second-order field equations but use the space-time metric as one of the
 fundamental fields. In a comparison with 2-dimensional Horndeski theories
 \cite{DilatonHorndeski}, the field $e_1$ here would be interpreted as a
 scalar or dilaton fields.

\subsection{Anisotropic models}

For anisotropic spatially homogeneous solutions of Kantowski--Sachs type, we
 assume that $q_{xx}^{\rm em}=a_1(t)^2$ and $e_1=a_2(t)^2$ are functions only
 of proper time, such that $N=1$. Solving (\ref{qeffpert}) for $e_2$, we obtain
 \begin{equation}
   e_2=\frac{a_1a_2\sqrt{f}}{\sqrt{1+\bar{\lambda}^2\dot{a}_2^2/f}}\approx
   a_1a_2\sqrt{f}\left(1-\frac{1}{2}\bar{\lambda}^2\frac{\dot{a}_2^2}{f}\right)\,.
 \end{equation}
 For $f=1$ and $\alpha_2=1$, the Hamiltonian constraint then takes the form
 \begin{eqnarray}
   \frac{T[1]}{\ell_0a_1a_2^2}&=& -\frac{1}{2}f_0-
                            \frac{\dot{a}_1}{a_1}\frac{\dot{a}_2}{a_2}-
                            \frac{1}{2} \frac{\dot{a}_2^2}{a_2^2}\\
   && +\bar{\lambda}^2a_2^2 \left(\frac{1}{4}f_0\frac{\dot{a}_2^2}{a_2^2}-
      \frac{3}{2} \frac{\dot{a}_1}{a_1} \frac{\dot{a}_2^3}{a_2^3}+
      \frac{3}{4}\frac{\dot{a}_2^4}{a_2^4}+ \frac{\dot{a}_2^2}{a_2^2}
      \left(\frac{\dot{a}_2}{a_2}\right)^{\bullet}\right)\nonumber
 \end{eqnarray}
where we divided by a volume factor $V_0/(4\pi)=\ell_0 a_1a_2^2$ with the
coordinate length $\ell_0=\int{\rm d}x$ of a radial interval used to specify
the homogeneous geometry. Once the constraint is imposed, any dependence on
$\ell_0$ disappears.

This equation, which amounts to an anisotropic version of the
  Friedmann equation when the constraint $T[1]=0$ is imposed, shows several
  crucial features: First, there may be higher-derivative couplings (through
  the last term) for the coefficient $a_2$ of an emergent metric, even though
  the underlying canonical equations of motion for $e_1$, $e_2$ and their
  momenta are first-order. Secondly, if there is a cosmological constant,
  which would contribute a constant term to $f_0$, it contributes to
  $\bar{\lambda}$-corrections through a coupling term to the expansion
  rates. Finally, $\bar{\lambda}$-corrections necessarily depend on the size
  $a_2$ of the spherical orbits through the coefficient
  $\bar{\lambda}^2a_2^2$. In homogeneous models of loop quantum cosmology, an
  attempt has often been made to avoid this feature by using a non-constant
  $\lambda\propto a_2^{-1}$ instead of a constant $\bar{\lambda}$ (translated
  to the present notation) but, as shown by emergent modified gravity, in
  covariant theories this dependence can only be obtained by a canonical
  transformation from the models studied here, at the expense of introducing
  extra terms in the constraint that again lead to a dependence of
  $\lambda$-corrections on the size $a_2$. Possible covariant corrections
  share this feature with corrections from quantum fluctuation terms, in which
  case the dependence can be interpreted as a gravitational analog of the
  Casimir effect \cite{Infrared}.

  If we define expansion and shear through
 \begin{equation}
 \theta=\frac{\dot{a}_1}{a_1}+2\frac{\dot{a}_2}{a_2}\quad,\quad
 \sigma=\frac{2}{3}\left(\frac{\dot{a}_1}{a_1}-\frac{\dot{a}_2}{a_2}\right)
\end{equation}
the constraint takes the form
\begin{eqnarray}
\frac{T[1]}{\ell_0a_1a_2^2} &=& -\frac{1}{2}f_0-
                          \frac{1}{6}\theta^2-\frac{3}{8}\sigma^2+
\bar{\lambda}^2a_2^2\left(\frac{1}{4}f_0\left(\frac{1}{3}\theta-\frac{1}{2}\sigma\right)^2\right.\\
    &&\qquad          
- \frac{3}{4}\left(\frac{1}{81}\theta^4+\frac{1}{27}\theta^3\sigma-
  \frac{1}{3}\theta^2\sigma^2+
          \frac{1}{4}\theta\sigma^3-\frac{5}{16}\sigma^4\right)\nonumber\\
  &&\qquad\left.+
\left(\frac{1}{9}\theta^2-\frac{1}{3}\theta\sigma+ \frac{1}{4}\sigma^2\right)
\left(\frac{1}{3}\dot{\theta}-\frac{1}{2}\dot{\sigma}\right)\right)\,. \nonumber
\end{eqnarray}
A Friedmann-type equation derived from these models therefore has additional
coupling terms between expansion and shear, as well as their time derivatives,
but only in $\bar{\lambda}$-corrections. If $\alpha_2\not=1$, an additional
term
\begin{equation}
(\alpha_2-1)\frac{\dot{a}_2^2}{a_2^2}=
(\alpha_2-1)\left(\frac{1}{8}\theta^2-\frac{1}{3}\theta\sigma+
  \frac{1}{4}\sigma^2\right)
\end{equation}
appears, which contributes another expansion-shear coupling independent of
$\bar{\lambda}$. A larger set of additional terms is implied if $f\not=1$ is
non-constant, since this function then changes the relationship between $e_2$
and $a_1$ as well as $a_2$. A large number of possible modifications of
Friedmann-type equations can therefore be obtained in covariant form.

\subsection{MONDified gravity}

The identification of the new concept of a well-defined emergent metric allows
us to look for new physical effects. The prime example found so far
\cite{HigherMOND} is an application to MOND (MOdified Newtonian Dynamics,
\cite{MOND1,MOND2}) which we briefly review here for the sake of
completeness. We have to solve field equations implied by the canonical
theory, in particular the constraints $T(N)=0$ and $S(w)=0$. We may also pick
conditions on some of the fields in order to choose a specific set of
solutions, such as static ones in which $k_1=0$ and $k_2=0$. The higher-order
modifications of the $k_2$-dependence that may have been introduced in $T(N)$
then disappear, at least in simple cases. (There are parameter choices in the
higher-order modifications that imply additional non-$k_2$-dependent terms in
$T(N)$, but we will not consider them here.)  Moreover, for solutions of the
constraints (as opposed to their off-shell gauge behavior that is important
for demonstrating general covariance of the modified theories) the linear
transformation (\ref{linear}) does not make a difference. We therefore obtain
the classical static solutions for the non-zero $e_1$ and $e_2$ as well as
$N$, just as they appear in the Schwarzschild line element. However, these
$e_1$ and $e_2$ now have to be inserted in the emergent line element
\begin{equation} \label{dsem}
    {\rm d}s_{\rm em}^2= -N(t,x)^2{\rm d}t^2+ q_{xx}^{\rm em}(e_1,e_2,k_2) ({\rm
      d}x+w(t,x){\rm d}t)^2+ e_1(t,x) {\rm d}\Omega^2
\end{equation}
with radial metric component (\ref{qeff2}), implying new and characteristic
physical effects.

The emergent metric directly appears in the normalization condition
$||p^2||=-m^2$ for the 4-momentum of a test particle of mass $m$ moving in our
space-time, from which an effective potential can be derived as usual.  Using
the standard conserved quantities $L$ (angular momentum) and $E$ (energy)
along a timelike geodesic, normalization implies
\begin{equation}
    -1 =
    q^{\rm em}_{xx} \left(\frac{{\rm d}x}{{\rm d}\tau}\right)^2
    + \frac{L^2}{m^2x^2}
    - \frac{E^2}{m^2N^2}
\end{equation}
with the lapse function $N$ of the space-time metric. This equation can be
rewritten as an energy-balance law,
\begin{equation}
    0 =
    \frac{1}{2} m \left(\frac{{\rm d}x}{{\rm d}\tau}\right)^2
    + \frac{q_{\rm em}^{x x}}{2} \left( m
    + \frac{L^2}{mx^2}
    - \frac{E^2}{mN^2} \right)
\end{equation}
with a kinetic radial energy and an effective potential.  Classsically, with
(\ref{ds}) in Schwarzschild form and $q^{xx}=e_1/e_2^2=1-2M/x$ with the
black-hole mass $M$, the $x$-dependence of the effective potential directly
implies Newton's potential

In an emergent metric of the form (\ref{qeff}), this potential may be
multiplied by higher-order terms in $1/x$ because
$\frac{1}{4}e_2^{-2} (\partial e_1/\partial x)^2\approx 1-2M/x$, depending on
the function $\lambda(e_1)$. Such terms would be relevant only at small radii
close to the Schwarzschild radius. The more general version (\ref{qeff2}),
however, contains a different function, $f(e_1)$, that implies additive
corrections to Newton's potential. In particular, it is possible to find
covariant modifications of spherically symmetric general relativity in which
$f(e_1)$ has a logarithmic contribution. Fundamentally, this is exactly what
is expected from fluctuation or renormalization effects in a quantum version
of gravity, as in the explicit example of \cite{SphSymmMoments}. Such terms
would be relevant on intermediate scales far from the horizon of a black hole,
for instance in the entire matter distribution around a galactic black
hole. In particular, exploiting the free parameters in modification functions
such as $f(e_1)$, it is not difficult to derive MOND-like effects
\cite{MOND1,MOND2} in the generally covariant setting of emergent modified
gravity, based on the emergent metric (\ref{qeff2}), in a fully covariant
manner and without the need for extra fields \cite{HigherMOND}.

A final observation demonstrates how the new effects of
  emergent modified gravity are intimately related to space-time structure. A
  specific example of a modification function $f(e_1)$ that leads to MOND-like
  effects is given by $f(e_1)=1-\bar{\lambda}^2 \log(e_1/c_0)$ with a new constant
  $c_0$ that could, for instance, result from renormalization. Using this
  function in (\ref{qeff2}) would imply a logarithmic contribution to an
  effective Newton potential. However, the right-hand side of (\ref{qeff2}) is
  then no longer positive for large $e_1$ or large radii, depending on the
  value of $c_0$. Unless the function $f(e_1)$ turns over to a different,
  non-logarithmic behavior at large radii, such models therefore describe
  large-scale signature change together with intermediate MOND effects. (See
  \cite{EmergentSig} for a detailed analysis of a related example.)
  Signature change of this form can be compatible with observations provided
  $c_0$ is sufficiently large, but it does affect the global space-time model
  implied by a specific version of emergent modified gravity.

\section{Conclusions}

At present, possible modifications in emergent modified gravity appear to be
far from unique, unless a specific quantum-gravity derivation is used. This
situation is comparable to the large class of higher-curvature or other
modified-gravity theories, that must be further restricted by fundamental
considerations or phenomenology. Just like the well-known cases, our new
theories produce computable theories in which many different effects can be
derived by the usual space-time analysis, using the existence of a
generally covariant emergent line element.

So far, MOND-like scenarios have been worked out in some detail, which may be
used to derive further constraints on the free functions.  In addition to the
effective potential, standard methods produce, for instance, results about
lightlike geodesics and small corrections to deflection angles. Combining
several of these effects, free parameters can be constrained by a detailed
analysis. For now, it is important to see that it is possible to obtain
MOND-like intermediate-scale effects in a generally covariant theory without
introducing new degrees of freedom, as required in other examples
\cite{TEVES,TEVES2}. Gravitational waves as perturbations around our emergent
background line elements may also be used as alternatives to higher-curvature
theories unburdened by the usual instabilities from higher-derivative terms,
providing new options to compare general relativity with other theories in
strong-field regimes. New effects of emergent modified gravity are
characterized by the form of the emergent line element, in which, even in a
static Schwarzschild-like background solution without extra fields, the
${\rm d}x^2$ and ${\rm d}t^2$ terms in the line element (\ref{ds}) are
modified in different ways, the former according to (\ref{qeff2}).

Finally, there is an interesting connection with quantum-gravity effects such
as higher-order contributions of extrinsic curvature to the Hamiltonian
constraint because the same mechanism of modified gravity that may give rise
to MOND-like effects through (\ref{qeff2}) also shows how obstructions to
general covariance found previously \cite{EffLine} in models of canonical
quantum gravity may be circumvented within the general class of modified
theories described by Hamiltonian constraints of the form (\ref{Hgen}).  Even
though MOND-like and quantum gravity phenomena usually happen on vastly
different scales, they may all benefit from a deeper understanding of
hypersurface deformation structures.

An important question is whether emergent modified gravity can be extended
non-trivially beyond vacuum spherical symmetry. Extensions to various matter
couplings such as scalar fields \cite{SphSymmMinCoup,EmergentScalar} and
perfect fluids \cite{EmergentFluid} within spherical symmetry have already
been found. Work in progress shows that an extension beyond spherical symmetry
is possible for cylindrically symmetric models, which are more general than
spherically symmetric ones in that they allow local gravitational degrees of
freedom. It is therefore possible to describe gravitational waves in this
setting. Moreover, the underlying equations used in \cite{HigherCov} to derive
modified Hamiltonian constraints and the emergent metric are based on
properties of hypersurface deformations. The general equations are available
without symmetry restrictions, but it remains to be seen whether they have
interesting solutions.

\section*{Acknowledgements}

This work was supported in part by NSF grant PHY-2206591.

\section*{References}

%\bibliographystyle{../preprint}
%\bibliography{../Bib/QuantGra,../Bib/Tunneling}

\begin{thebibliography}{10}

\bibitem{DiracHamGR}
P.~A.~M.\ Dirac,
\newblock The theory of gravitation in Hamiltonian form,
\newblock {\em Proc.\ Roy.\ Soc.\ A} 246 (1958) 333--343

\bibitem{Katz}
J.\ Katz,
\newblock Les crochets de Poisson des contraintes du champ gravitationne,
\newblock {\em Comptes Rendus Acad.\ Sci.\ Paris} 254 (1962) 1386--1387

\bibitem{ADM}
R.\ Arnowitt, S.\ Deser, and C.~W.\ Misner,
\newblock The Dynamics of General Relativity, In L.\ Witten, editor, {\em
  Gravitation: An Introduction to Current Research},
\newblock Wiley, New York, 1962,
\newblock Reprinted in \cite{ADMRe}

\bibitem{Regained}
S.~A.\ Hojman, K.\ Kucha\v{r}, and C.\ Teitelboim,
\newblock Geometrodynamics Regained,
\newblock {\em Ann.\ Phys.\ (New York)} 96 (1976) 88--135

\bibitem{EmergentSig}
M.\ Bojowald, E.~I.\ Duque, and D.\ Hartmann,
\newblock A new type of large-scale signature change in emergent modified
  gravity,
\newblock {\em Phys.\ Rev.\ D} 109 (2024) 084001, [arXiv:2312.09217]

\bibitem{SphSymmEff}
A.\ Alonso-Bardaj\'{\i}, D.\ Brizuela, and R.\ Vera,
\newblock An effective model for the quantum Schwarzschild black hole,
\newblock {\em Phys.\ Lett.\ B} 829 (2022) 137075, [arXiv:2112.12110]

\bibitem{SphSymmEff2}
A.\ Alonso-Bardaj\'{\i}, D.\ Brizuela, and R.\ Vera,
\newblock Nonsingular spherically symmetric black-hole model with holonomy
  corrections,
\newblock {\em Phys.\ Rev.\ D} 106 (2022) 024035, [arXiv:2205.02098]

\bibitem{ConsRinehart}
C.\ Blohmann, M.\ Schiavina, and A.\ Weinstein,
\newblock A Lie-Rinehart algebra in general relativity, [arXiv:2201.02883]

\bibitem{SphSymmCov}
M.\ Bojowald, S.\ Brahma, and J.~D.\ Reyes,
\newblock Covariance in models of loop quantum gravity: Spherical symmetry,
\newblock {\em Phys.\ Rev.\ D} 92 (2015) 045043, [arXiv:1507.00329]

\bibitem{GowdyCov}
M.\ Bojowald and S.\ Brahma,
\newblock Covariance in models of loop quantum gravity: Gowdy systems,
\newblock {\em Phys.\ Rev.\ D} 92 (2015) 065002, [arXiv:1507.00679]

\bibitem{EffLine}
M.\ Bojowald, S.\ Brahma, and D.-H.\ Yeom,
\newblock Effective line elements and black-hole models in canonical (loop)
  quantum gravity,
\newblock {\em Phys.\ Rev.\ D} 98 (2018) 046015, [arXiv:1803.01119]

\bibitem{SphSymmMatter}
A.\ Alonso-Bardaj\'{\i} and D.\ Brizuela,
\newblock Holonomy and inverse-triad corrections in spherical models coupled to
  matter,
\newblock {\em Eur.\ Phys.\ J.\ C} 81 (2021) 283, [arXiv:2010.14437]

\bibitem{SphSymmMatter2}
A.\ Alonso-Bardaj\'{\i} and D.\ Brizuela,
\newblock Anomaly-free deformations of spherical general relativity coupled to
  matter,
\newblock {\em Phys.\ Rev.\ D} 104 (2021) 084064, [arXiv:2106.07595]

\bibitem{Absorb}
R.\ Tibrewala,
\newblock Inhomogeneities, loop quantum gravity corrections, constraint algebra
  and general covariance,
\newblock {\em Class.\ Quantum Grav.} 31 (2014) 055010, [arXiv:1311.1297]

\bibitem{SphSymmHam}
M.\ Bojowald and R.\ Swiderski,
\newblock Spherically Symmetric Quantum Geometry: Hamiltonian Constraint,
\newblock {\em Class.\ Quantum Grav.} 23 (2006) 2129--2154, [gr-qc/0511108]

\bibitem{HigherCov}
M.\ Bojowald and E.~I.\ Duque,
\newblock Emergent modified gravity: Covariance regained,
\newblock {\em Phys.\ Rev.\ D} 108 (2023) 084066, [arXiv:2310.06798]

\bibitem{NonSingdeSitter}
V.\ Mukhanov and R.\ Brandenberger,
\newblock A nonsingular universe,
\newblock {\em Phys.\ Rev.\ Lett.} 68 (1992) 1969--1972

\bibitem{NonCovPol}
M.\ Bojowald,
\newblock Non-covariance of ``covariant polymerization'' in models of loop
  quantum gravity,
\newblock {\em Phys.\ Rev.\ D} 103 (2021) 126025, [arXiv:2102.11130]

\bibitem{CovPol}
F.\ Ben\'{\i}tez, R.\ Gambini, and J.\ Pullin,
\newblock A covariant polymerized scalar field in loop quantum gravity,
\newblock {\em Universe} 8 (2022) 526, [arXiv:2102.09501]

\bibitem{JR}
J.~D.\ Reyes,
\newblock {\em Spherically Symmetric Loop Quantum Gravity: Connections to
  2-Dimensional Models and Applications to Gravitational Collapse},
\newblock PhD thesis, The Pennsylvania State University, 2009

\bibitem{Horndeski}
G.~W.\ Horndeski,
\newblock Second-order scalar-tensor field equations in a four-dimensional
  space,
\newblock {\em Int.\ J.\ Theor.\ Phys.} 10 (1974) 363--384

\bibitem{HorndeskiRev}
T.\ Kobayashi,
\newblock Horndeski theory and beyond: a review,
\newblock {\em Rept.\ Prog.\ Phys.} 82 (2019) 086901, [arXiv:1901.07183]

\bibitem{DHOST}
D.\ Langlois and K.\ Noui,
\newblock Degenerate higher derivative theories beyond Horndeski: evading the
  Ostrogradski instability,
\newblock {\em JCAP} 02 (2016) 034, [arXiv:1510.06930]

\bibitem{DilatonHorndeski}
K.\ Takahashi and T.\ Kobayashi,
\newblock Generalized 2D dilaton gravity and KGB,
\newblock {\em Class.\ Quant.\ Grav.} 36 (2019) 095003, [arXiv:1812.08847]

\bibitem{Action}
M.\ Bojowald and G.~M.\ Paily,
\newblock Deformed General Relativity and Effective Actions from Loop Quantum
  Gravity,
\newblock {\em Phys.\ Rev.\ D} 86 (2012) 104018, [arXiv:1112.1899]

\bibitem{Horava}
P.\ Ho\v{r}ava,
\newblock Quantum gravity at a Lifshitz point,
\newblock {\em Phys.\ Rev.\ D} 79 (2009) 084008, [arXiv:0901.3775]

\bibitem{Infrared}
M.\ Bojowald,
\newblock The BKL scenario, infrared renormalization, and quantum cosmology,
\newblock {\em JCAP} 01 (2019) 026, [arXiv:1810.00238]

\bibitem{HigherMOND}
M.\ Bojowald and E.~I.\ Duque,
\newblock MONDified gravity,
\newblock {\em Phys.\ Lett.\ B} 847 (2023) 138279, [arXiv:2310.19894]

\bibitem{MOND1}
M.\ Milgrom,
\newblock A modification of the Newtonian dynamics-Implications for galaxies,
\newblock {\em Ap.\ J.} 270 (1983) 371--383

\bibitem{MOND2}
S.~S.\ McGaugh and W.\ De~Blok,
\newblock Testing the hypothesis of modified dynamics with low surface
  brightness galaxies and other evidence,
\newblock {\em Ap.\ J.} 499 (1998) 66

\bibitem{SphSymmMoments}
K.\ Berglund, M.\ Bojowald, M.\ D\'{\i}az, and G.\ Sims,
\newblock Quasiclassical solutions for static quantum black holes,
\newblock {\em Phys.\ Rev.\ D} 109 (2024) 024006, [arXiv:2012.07649]

\bibitem{TEVES}
J.~D.\ Bekenstein,
\newblock Relativistic gravitation theory for the modified Newtonian dynamics
  paradigm,
\newblock {\em Phys.\ Rev.\ D} 70 (2004) 083509, [astro-ph/0403694]

\bibitem{TEVES2}
J.~W.\ Moffat,
\newblock Scalar–tensor–vector gravity theory,
\newblock {\em JCAP} 2006 (2006) 004, [gr-qc/0506021]

\bibitem{SphSymmMinCoup}
A.\ Alonso-Bardaj\'{\i} and D.\ Brizuela,
\newblock Spacetime geometry from canonical spherical gravity,
  [arXiv:2310.12951]

\bibitem{EmergentScalar}
M.\ Bojowald and E.~I.\ Duque,
\newblock Emergent modified gravity coupled to scalar matter,
\newblock {\em Phys.\ Rev.\ D} 109 (2024) 084006, [arXiv:2311.10693]

\bibitem{EmergentFluid}
E.~I.\ Duque,
\newblock Emergent modified gravity: The perfect fluid and gravitational
  collapse,
\newblock {\em Phys.\ Rev.\ D} 109 (2024) 044014, [arXiv:2311.08616]

\bibitem{ADMRe}
R.\ Arnowitt, S.\ Deser, and C.~W.\ Misner,
\newblock The Dynamics of General Relativity,
\newblock {\em Gen.\ Rel.\ Grav.} 40 (2008) 1997--2027

\end{thebibliography}

\end{document}